\documentclass[amsmath,amssymb,aps,pra,reprint,10pt]{revtex4-1}
\usepackage{bm}
\usepackage{amsfonts}
\usepackage{amssymb}
\usepackage{graphicx}

\begin{document}

\title{Simple model of self-supported deformed states of isolated atoms}
\author{Iwo Bialynicki-Birula}\email{birula@cft.edu.pl}
\affiliation{Center for Theoretical Physics, Polish Academy of Sciences\\
Al. Lotnik\'ow 32/46, 02-668 Warsaw, Poland}
\author{Zofia Bialynicka-Birula}\affiliation{Institute of Physics, Polish Academy of Sciences\\
Al. Lotnik\'ow 32/46, 02-668 Warsaw, Poland}

\begin{abstract}
We propose a simple three-body model of an atom in which one electron on a circular Rydberg orbit is treated as an independent particle and the remaining core electrons are collectively described as a single object. Within this model we predict the existence of stable deformed states of atoms. The deformation is generated by a bootstrap mechanism. The atomic core is polarized by the excited electron and the induced dipole moment keeps this electron localized. The deformed stable states of the atom are similar to the Trojan states observed in recent experiments. However, in the present case the breaking of the rotational symmetry does not require the presence of external fields.
\end{abstract}
\maketitle

\section{Introduction}

The prediction in Ref.~\cite{bbke} of nonspreading wave packets describing Rydberg electrons driven by a circularly polarized wave was recently fully confirmed in experiments \cite{mgg,metal}. The dynamics of these nonstationary but stable electronic states resembles the dynamics of Trojan asteroids in the Sun-Jupiter system. The Trojan states of atoms are deformed due to the action of the driving field. In this article we propose a simple model to study a possible existence of deformed states of multielectron atoms not subjected to external forces. The deformation is caused by the interaction of a Rydberg electron with collectively described core electrons. The interaction of the Rydberg electron with the core has been studied by Clark and Greene \cite {cg} who described this interaction by a phenomenological coupling of the angular momentum of the electron to the angular momentum of the ion core. They were able to account successfully for subtle details of spectroscopic data but the deformation of the atoms has not been the subject of their interest. In our model we keep the pure Coulombic coupling of the Rydberg electron to the core but we describe the interaction of the core with nucleus by harmonic forces. Recently Kalinski {\em et al.} \cite{k4} have shown in the simplest case of the helium atom that due to the core polarizability deformed localized states can exist in isolated atoms. The mechanism responsible for the deformation was clearly identified in Ref.~\cite{k4}, where we can find the following description: ``The oscillations of the inner charge generates a stabilizing field for the outer electron''. The analysis presented in Ref.~\cite{k4} was specifically designed for helium and its extension to many-electron atoms would be very complicated. We propose a simple model that captures the essential features of such special states of atoms. In our model one highly excited electron acting on {\em all} inner electrons creates a dipole moment and the induced dipole localizes the outer electron on its orbit.

Our aim is to give an intuitive picture of deformed atomic states that explains the nature of these states and could enable one to pinpoint their characteristics without the use of more elaborate methods. The purpose of the present article is, therefore, to show that the mechanism proposed in Ref.~\cite{k4} can produce deformed states of many-electron atoms. There is, however, a crucial difference between our work and that of Ref.~\cite{k4}. In our case the mutual interaction of the core electrons and the outer electron leads to a {\em stable} configuration. The stability analysis in our model can be carried out analytically. The correlations of the electron and the core coordinates can be clearly exhibited. The strong point of our analysis is that many steps can be performed analytically.

There is an essential difference between our stable electronic states and the ones supported by an external field. Since the dynamics of an {\em isolated atom} is governed by a rotationally symmetric Hamiltonian, the deformed states can only occur due to {\em broken} rotational symmetry. This mechanism is well known in nuclear physics where the breaking of rotational symmetry leads to deformed states of nuclei \cite{rs,c}. The analogy between the deformed states of nuclei and possible deformed states of atoms was pointed out by Iwai \cite{iwai}.

\section{The model}

To describe intricate dynamics that leads to a deformed atom, we introduce a greatly simplified model of a multielectron atom. Within this model we can identify atomic states in which the repulsion between one excited electron and the core causes the deformation. This lonely electron will be described as usual but the remaining core electrons occupying orbits close to the nucleus will be replaced by a {\em single effective particle} endowed with the mass and the charge of all the core electrons. The interaction of this effective particle with the far away electron will be kept Coulombic while the interaction with nucleus will be modeled by a harmonic force with a spring constant reflecting the polarizability of the core. The description of the strongly bound electrons in terms of a harmonic force and the corresponding oscillator strength parameters has been successfully used in the early days of quantum mechanics. Here we push this description further by applying the combined oscillator model to all core electrons. Our model cannot give a precise description of the deformed atom but owing to its simplicity gives a clear picture of the mechanism that is responsible for the symmetry breaking and for the stability of deformed states.

The Hamiltonian in our model of the atom has the following form:
\begin{align}\label{ham}
H&=\frac{{\bm p}_n^2}{2m_n}+\frac{{\bm p}_e^2}{2m_e}-\frac{Zq^2}{|{\bm r}_e-{\bm r}_n|}\nonumber\\
&+\frac{{\bm p}_c^2}{2m_c}+\frac{k}{2}({\bm r}_c-{\bm r}_n)^2+\frac{(Z-1)q^2}{|{\bm r}_e-{\bm r}_c|},
\end{align}
where $q^2=e^2/4\pi\epsilon_0$. The first part of this Hamiltonian describes the kinetic energies of the nucleus and the singled out electron, and also their mutual Coulomb interaction. The second part describes the kinetic energy of the atomic core, its harmonic interaction with the nucleus, and the Coulomb interaction of the core with the electron. Our model may seem to be best suited to describe the alkali atoms for which the distinction between the core and the valence electron is well defined. However, the polarizability of alkali ions is very small. Therefore, as an example, we have chosen magnesium, whose ion has the same configuration of electrons as sodium.

Translational invariance allows for the separation of the center of mass motion. This is achieved by the following canonical transformation:
\begin{subequations}\label{cm}
\begin{align}
{\bm r}_e-{\bm r}_n={\bm r},\quad{\bm r}_c-{\bm r_n}&={\bm R},\\
\frac{m_e{\bm r}_e+m_c{\bm r}_c+m_n{\bm r}_n}{M}&={\bm r}_{\rm cm},\\
\frac{(m_n+m_c){\bm p}_e-m_e({\bm p}_n+{\bm p}_c)}{M}&={\bm p},\\
\frac{(m_e+m_n){\bm p}_c-m_c({\bm p}_e+{\bm p}_n)}{M}&={\bm P},\\
{\bm p}_e+{\bm p}_c+{\bm p}_n&={\bm p}_{\rm cm},
\end{align}
\end{subequations}
where $M=m_e+m_c+m_n$. The inverse transformation reads:
\begin{subequations}\label{icm}
\begin{align}
{\bm r}_e&={\bm r}_{\rm cm}+\frac{(m_c+m_n){\bm r}-m_c{\bm R}}{M},\\
{\bm r}_c&={\bm r}_{\rm cm}+\frac{(m_e+m_n){\bm R}-m_e{\bm r}}{M},\\
{\bm r}_n&={\bm r}_{\rm cm}-\frac{m_e{\bm r}+m_c{\bm R}}{M},\\
{\bm p}_e&=\frac{m_e{\bm p}_{\rm cm}}{M}+{\bm p},\\
{\bm p}_c&=\frac{m_c{\bm p}_{\rm cm}}{M}+{\bm P},\\
{\bm p}_n&=\frac{m_n{\bm p}_{\rm cm}}{M}-{\bm p}-{\bm P}.
\end{align}
\end{subequations}
The Hamiltonian (\ref{ham}) expressed in terms of the new variables split into the sum $H=H_{\rm cm}+H_{\rm rel}$. Choosing the coordinate system in which the center of mass is at rest, we may drop $H_{\rm cm}={\bm p}_{\rm cm}^2/2M$ and keep only the Hamiltonian of the relative motion $H_{\rm rel}$,
\begin{align}\label{ham1}
H_{\rm rel}=\frac{\bm p^2}{2\mu_e}+\frac{\bm P^2}{2\mu_c}+\frac{{\bm p}\!\cdot\!{\bm P}}{m_n}
+\frac{k}{2}{\bm R}^2+\frac{(Z-1)q^2}{|{\bm r}-{\bm R}|}-\frac{Zq^2}{|{\bm r}|},
\end{align}
where $\mu_e$ and $\mu_c$ are the reduced masses of the electron-nucleus and the core-nucleus systems,
\begin{align}\label{mu}
\mu_e=\frac{m_e\,m_n}{m_e+m_n},\quad\mu_c=\frac{m_c\,m_n}{m_c+m_n}.
\end{align}

The system described by the Hamiltonian $H_{\rm rel}$ has a stable stationary configuration. This state in the rotating frame becomes an easily identifiable state of {\em static equilibrium}. In the frame rotating with the angular velocity vector ${\bm\omega}$ the Hamiltonian acquires an additional term $-{\bm\omega}\!\cdot\!({\bm r}\times{\bm p}+{\bm R}\times{\bm P})$ describing the inertial forces.

From now on we shall express all quantities in atomic units. The Hamiltonian in the rotating frame ${\cal H}$ has the form,
\begin{align}\label{hamr}
{\cal H}&=\frac{{\bm p}^2}{2\mu_e}+\frac{{\bm P}^2}{2\mu_c}
+\frac{{\bm p}\!\cdot\!{\bm P}}{m_n}-\frac{Z}{|{\bm r}|}+\frac{Z-1}{|{\bm r}-{\bm R}|}\nonumber\\
&+\frac{k}{2}{\bm R}^2-{\bm\omega}\!\cdot\!({\bm r}\times{\bm p}+{\bm R}\times{\bm P}).
\end{align}

The canonical equations of motion generated by the Hamiltonian ${\cal H}$ are:
\begin{subequations}\label{eqm}
\begin{align}
\frac{d{\bm r}}{dt}&=\frac{{\bm p}}{\mu_e}+\frac{{\bm P}}{m_n}-{\bm\omega}\times{\bm r},\label{eqm1}\\
\frac{d{\bm R}}{dt}&=\frac{{\bm P}}{\mu_c}+\frac{{\bm p}}{m_n}-{\bm\omega}\times{\bm R},\label{eqm2}\\
\frac{d{\bm p}}{dt}&=-\frac{Z{\bm r}}{|{\bm r}|^3}+\frac{(Z-1)({\bm r}-{\bm R})}
{|{\bm r}-{\bm R}|^3}-{\bm\omega}\times{\bm p},\label{eqm3}\\
\frac{d{\bm P}}{dt}&=-k{\bm R}-\frac{(Z-1)({\bm r}-{\bm R})}{|{\bm r}-{\bm R}|^3}
-{\bm\omega}\times{\bm P}.\label{eqm4}
\end{align}
\end{subequations}
For a given atom all masses, $Z$, and $k$ are fixed and the dynamics of the system is controlled by just one parameter $\omega$. This system has two constants of motion: the total energy and the total angular momentum.

\section{Equilibrium configuration}

The system is in equilibrium when all time derivatives vanish---all bodies are at rest. The mere existence of an equilibrium configuration in rotating frames is not a special property of our model. Such configurations exist for all atoms with {\em pure Coulombic forces}. They correspond to the regular, equidistant positions of electrons. However, these equilibria are highly unstable. In our model the stability is achieved by a collective description of the core electrons.

In the equilibrium configuration described by Eqs.~(\ref{eqm1})--(\ref{eqm4}) the excited valence electron, when viewed from the laboratory frame, moves on a circular Kepler orbit with frequency $\omega$ around the center of mass and the core particle and the nucleus move on much smaller circular orbits. After the elimination of momenta, we obtain the following two equations for the equilibrium position vectors:
\begin{subequations}\label{eqe}
\begin{align}
\frac{Z{\bm r}_0}{|{\bm r}_0|^3}
&=\frac{(Z-1)({\bm r}_0-{\bm R}_0)}{|{\bm r}_0-{\bm R}_0|^3}\nonumber\\
&+\omega^2\!\left(\frac{m_e m_c+m_e m_n}{M}{\bm r}_0-\frac{m_e m_c}{M}{\bm R}_0\right)_{\!\!\perp},\label{eqe1}\\
k{\bm R}_0&=\frac{(Z-1)({\bm R}_0-{\bm r}_0)}{|{\bm r}_0-{\bm R}_0|^3}\nonumber\\
&+\omega^2\!\left(\frac{m_e m_c+m_c m_n}{M}{\bm R}_0
-\frac{m_e m_c}{M}{\bm r}_0\right)_{\!\!\perp},\label{eqe2}
\end{align}
\end{subequations}
where $_\perp$ denotes the component perpendicular to $\bm\omega$. These equilibrium conditions express simply the balance of forces acting on each particle. It follows from these equations that ${\bm r}_0$ and ${\bm R}_0$ are aligned and lie in the plane perpendicular to $\bm\omega$. Since the core is repelled by the electron while the nucleus is attracted, these two vectors point in opposite directions. We shall choose the $x$ axis in the direction of ${\bm r}_0$ and denote by $x_1$ and $-x_2$ the $x$ components of ${\bm r}_0$ and ${\bm R}_0$. These parameters represent the distances from the electron to the nucleus and from the core to the nucleus. Eqs.~(\ref{eqe1})--(\ref{eqe2}) rewritten in terms of $x_1$ and $x_2$ form a set of two coupled third-order equations,
\begin{subequations}\label{eqa}
\begin{align}
\frac{Z}{x_1^2}&=kx_2+\omega^2\frac{m_e m_nx_1-m_c m_nx_2}{M},\label{eqa1}\\
\frac{(Z-1)}{(x_1+x_2)^2}&=kx_2-\omega^2\frac{(m_e m_c+m_c m_n)x_2+m_e m_cx_1}{M}.\label{eqa2}
\end{align}
\end{subequations}
The first equation is obtained as a difference of equations (\ref{eqe1}) and (\ref{eqe2}).

Let us denote by $\delta$ the relative distance from the core to the nucleus,
\begin{align}\label{prop}
\delta=x_2/x_1.
\end{align}
After dividing Eq.~(\ref{eqa1}) by Eq.~(\ref{eqa2}) we obtain
\begin{align}\label{eqz}
\frac{Z(1+\delta)^2}{Z-1}=\frac{kM\delta+\omega^2(m_e m_n-m_c m_n\delta)}
{kM\delta-\omega^2\left[m_e m_c+(m_c m_n+m_e m_c)\delta\right]},
\end{align}
and this leads to a single third-order equation for $\delta$. Full solution of the equilibrium conditions can be obtained in the general case. Here we present only the results obtained in the limit of an infinite nucleus mass. We could proceed without making this approximation, but in the case of a many electron atom the results would differ by a small fraction of a percent. In this limit the equations for $x_1$, $x_2$, and $\delta$ read
\begin{align}
\frac{Z}{x_1^2}&=\omega^2x_1+[k-(Z-1)\omega^2]x_2,\label{eqb1}\\
\frac{Z-1}{(x_1+x_2)^2}&=[k-(Z-1)\omega^2]x_2,\label{eqb2}\\
\delta^3+2\delta^2+\frac{\delta}{Z}&=\frac{(1-1/Z)}{k/\omega^2-Z+1},\label{eqb3}
\end{align}
where we explicitly used the fact that the core mass measured in units of the electron mass is $Z-1$. It follows from Eq.~(\ref{eqb3}) that the value of $\delta$ tends to zero together with $\omega$,
\begin{align}\label{dw}
\delta\approx\frac{Z-1}{k}\omega^2.
\end{align}

Since $\delta$ must be positive, Eq.~(\ref{eqb3}) imposes an upper bound on $\omega$, namely $(Z-1)\omega^2<k$. This bound simply means that when the rotation is too fast the centrifugal force will overcome the spring tension and the core will fly away. On physical grounds this upper bound is a bit too large. When the distance from the core to the nucleus becomes equal to the distance from the electron to the nucleus, our model becomes clearly inapplicable. This, according to Eq.~(\ref{eqb1}), leads to a slightly more stringent condition
\begin{align}\label{bound}
\omega^2<\frac{3Z+1}{(Z-1)(3Z+2)}k.
\end{align}

Equation (\ref{eqb3}) can be easily solved for $\omega$,
\begin{align}\label{del}
\omega(\delta,k,Z)=\left[\frac{\delta+2Z\delta^2+Z\delta^3}{(Z-1)(1+\delta+2Z\delta^2+Z\delta^3)}
k\right]^{1/2}.
\end{align}
Alternatively, it can also be solved for $\delta$,
\begin{align}\label{del1}
\delta(\omega,k,Z)
=\frac{\sqrt[3]{a\!+\!\sqrt{a^2-4b^3}}\!+\!\sqrt[3]{a\!-\!\sqrt{a^2-4b^3}}}{3\sqrt[3]{2}}-\frac{2}{3},
\end{align}
where
\begin{subequations}\label{ab}
\begin{align}
a&=27\frac{(1-1/Z)\omega^2}{k-(Z-1)\omega^2}+\frac{18}{Z}-16,\\
b&=4-3/Z.
\end{align}
\end{subequations}
Now, dividing both sides of Eq.~(\ref{eqb1}) by $x_1$ we obtain the distance $x_1$ from the electron to the nucleus as a function of $\omega,\,k$, and $Z$,
\begin{align}\label{x1}
x_1(\omega,k,Z)=\left[\frac{Z}{\omega^2+[k-(Z-1)\omega^2]\,\delta(\omega,k,Z)}\right]^{1/3}\!\!\!,
\end{align}
or as a function of $\delta,\,k$, and $Z$,
\begin{align}\label{x1d}
x_1(\delta,k,Z)=\left[\frac{(Z-1)(1+\delta+2Z\delta^2+Z\delta^3)}{\delta(1+\delta)^2k}\right]^{1/3}.
\end{align}
The distance $x_2$ from the core to the nucleus is obtained by multiplying $x_1$ by $\delta$. When $\omega\to 0$ we obtain
\begin{align}\label{kep}
x_1(\omega,k,Z)\approx \frac{1}{\omega^{2/3}}.
\end{align}
This result is in agreement with the classical formula for the radius of a high hydrogenic Rydberg orbit because for such orbits the only effect that the core has on the electron is the screening of the nuclear charge. In the equilibrium configuration, the $z$-component total angular momentum along $\bm\omega$ is
\begin{align}\label{am}
M_z=\omega\left[x_1(\omega,k,Z)^2+(Z-1)x_2(\omega,k,Z)^2\right].
\end{align}

\section{Physical interpretation of the equilibrium configuration}

The results of the previous section confirm our intuitive picture of the deformed atom. As might be expected, in the equilibrium configuration all three objects, the core, the nucleus, and the electron, lie on a straight line. The core and the electron are on the opposite sides of the nucleus. In planetary terminology, the equilibrium configuration in our model corresponds to the Lagrange point $L_3$---all bodies are aligned and the planets are on the opposite sides of the Sun. Unlike its planetary counterpart, however, our equilibrium configuration is stable (for sufficiently small $\omega$). In the next section we prove the stability of motion near the equilibrium by calculating the characteristic frequencies in the quadratic approximation to the Hamiltonian.

\begin{figure}
\centering
\includegraphics[scale=0.6]{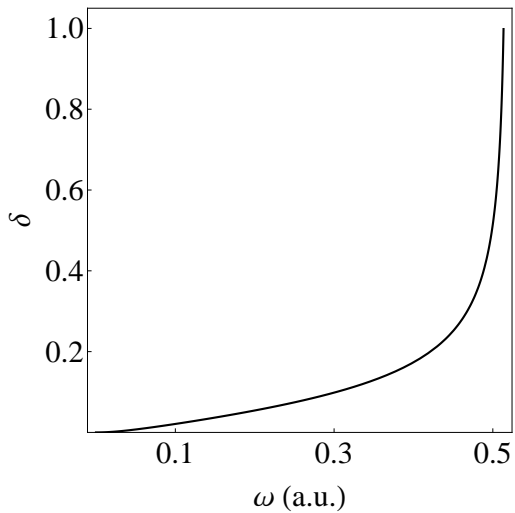}
\caption{The relative distance $\delta$ from the core to the nucleus as a function of $\omega$ for magnesium.}\label{fig1}
\end{figure}

\begin{figure}
\centering
\includegraphics[scale=0.6]{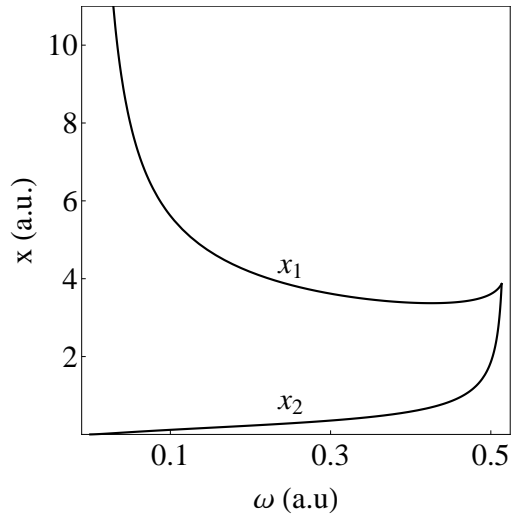}
\caption{The distances between the electron and the nucleus $x_1$ and between the core and the nucleus $x_2$ as a function of $\omega$ for magnesium. Distances are measured in atomic units.}\label{fig2}
\end{figure}

All calculations in this article will be done for the magnesium atom: $m_c=11\,m_e,\;m_n=43710\,m_e$, and $Z=12$. To complete this set of data we will need the value of the spring constant $k$. This can be determined from the polarizability. We can relate the parameter $k$ in our model to the polarizability $\alpha$, defined in the standard manner as the ratio of the induced electric dipole to the applied electric field. The induced dipole is $(Z-1)e {\bm R}$ and the electric field $E$ of the electron that creates this dipole by displacing the core away from the nucleus is
\begin{align}\label{el}
E=\frac{e}{4\pi\epsilon_0|{\bm r}-{\bm R}|^2}.
\end{align}
Using the values at the equilibrium, we arrive at the formula
\begin{align}\label{pol}
\tilde\alpha &=\frac{\alpha}{4\pi\epsilon_0}=(Z-1)|{\bm R}_0||{\bm r}_0-{\bm R}_0|^2\nonumber\\
&=(Z-1)|{\bm r}_0|^3\delta(1+\delta)^2.
\end{align}
The right-hand side is a function of the model parameters and $\omega$. In order to obtain the static polarizability we should take the limit when $\omega\to 0$ or $\delta\to 0$. Therefore, we may use the asymptotic formula,
\begin{align}\label{as}
|{\bm r}_e-{\bm r}_n|^3=x_1^3\approx \frac{Z-1}{k\delta},
\end{align}
obtained from Eq.~(\ref{x1d}) when $\delta\to 0$. In this limit we arrive at an approximate relation between the static polarizability and $k$,
\begin{align}\label{polk}
\tilde\alpha\approx\frac{(Z-1)^2e^2}{4\pi\epsilon_0k}.
\end{align}

This formula can also be obtained directly from the following observation. In the limit when $|{\bm R}|/|{\bm r}|\to 0$ and $m_n\to\infty$ the Hamiltonian (\ref{ham1}) describes two noninteracting systems: an electron in the Coulomb field and an oscillator with mass $(Z-1)m_e$ and the spring constant $k$. The displacement of an oscillator of charge $(Z-1)e$ under the action of the applied electric field $E$ is $x=(Z-1)eE/k$. Thus, the polarizability defined as the ratio of the induced electric dipole $(Z-1)e\,x$ to the electric field [Eq.~(\ref{el})] is given by the formula Eq.~(\ref{polk}).

Using the value of the rationalized static polarizability $\tilde\alpha$ of the magnesium ion \cite{tcn},
\begin{align}\label{polion}
\tilde\alpha=34.62\,{\rm a.u.}=5.13\times 10^{-30}{\rm m}^3,
\end{align}
we obtain the following estimate
\begin{align}\label{k}
k\approx 3.5\,{\rm a.u.}=5442\,{\rm kg\,s}^{-2}.
\end{align}
The characteristic energy of the core oscillations corresponding to this value is $\hbar\sqrt{k/m_c}=15.3\,{\rm eV}$, a fairly reasonable number. Having fixed the parameter $k$ of our model, we can proceed with detailed calculations.

All numerical calculations and figures in this article were done with the use of MATHEMATICA \cite{w}. The function $\delta(\omega,Z)$ for magnesium is depicted in Fig.~\ref{fig1}. The value of $\delta$---the relative distance from the core to the nucleus---is a good measure of the atomic deformation. In Fig.~\ref{fig2} we plotted the values of the equilibrium coordinates $x_1$ and $x_2$ expressed in Bohr radii $a_0$. At the value [Eq.~(\ref{bound})] of $\omega$ they become equal.

\section{Stability of a deformed state}

We will investigate the stability of the equilibrium state in our model by linearizing the equations of motion near the equilibrium configuration. This standard method worked very well in the studies of Trojan states \cite{bbke,bbbb1,bbbb}. We shall begin with the linearized equations of motion obtained from Eqs.~(\ref{eqm}) in the limit of infinite nucleus mass,
\begin{subequations}\label{eql}
\begin{align}
{\dot{\bm\alpha}}_1&={\bm\beta}_1-{\bm\omega}\times{\bm\alpha}_1,\label{eql1}\\
{\dot{\bm\beta}}_1&=(Z-1)\frac{{\bm\alpha}_1-{\bm\alpha}_2-3{\bm m}\left[{\bm m}\!\cdot\!({\bm\alpha}_1-{\bm\alpha}_2)\right]}
{(x_1+x_2)^3}\nonumber\\
&-Z\frac{{\bm\alpha}_1-3{\bm m}({\bm m}\!\cdot\!{\bm\alpha}_1)}{x_1^3}-{\bm\omega}\times{\bm\beta}_1,\label{eql2}\\
{\dot{\bm\alpha}}_2&=\frac{{\bm\beta}_2}{Z-1}-{\bm\omega}\times{\bm\alpha}_2,\label{eql3}\\
{\dot{\bm\beta}}_2&=-(Z-1)\frac{{\bm\alpha}_1-{\bm\alpha}_2-3{\bm m}\left[{\bm m}\!\cdot\!({\bm\alpha}_1-{\bm\alpha}_2)\right]}{(x_1+x_2)^3}\nonumber\\
&-k{\bm\alpha}_2-{\bm\omega}\times{\bm\beta}_2,\label{eql4}\\
{\dot{\zeta}}_1&=\sigma_1,\label{eql5}\\
{\dot{\sigma}}_1&=(Z-1)\frac{\zeta_1-\zeta_2}{(x_1+x_2)^3}
-Z\frac{\zeta_1}{x_1^3},\label{eql6}\\
{\dot{\zeta}}_2&=\frac{\sigma_2}{Z-1},\label{eql7}\\
{\dot{\sigma}}_2&=-k\zeta_2-(Z-1)\frac{\zeta_1-\zeta_2}{(x_1+x_2)^3},\label{eql8}
\end{align}
\end{subequations}
where the two-dimensional vectors ${\bm\alpha}_1,{\bm\beta}_1,{\bm\alpha}_2,{\bm\beta}_2$ describe small deviations from the equilibrium positions and momenta of both particles in the $xy$ plane, while $\zeta_1,\sigma_1,\zeta_2,\sigma_2$ describe the deviations along the $z$ axis. The unit vector parallel to the line passing through the equilibrium positions (in our case the $x$ axis) is denoted by  ${\bm{m}}$. Note that in this approximation the motion in the $z$ direction decouples from the motion in the $xy$ plane.

The right-hand side of the linearized equations of motion can be encoded in the following two matrices:
\begin{align}\label{eqmxy}
{\cal M}_{xy}=\left[\begin{array}{cccccccc}
0&\omega&1&0&0&0&0&0\\
-\omega&0&0&1&0&0&0&0\\
2a_1&0&0&\omega&2a_2&0&0&0\\
0&-a_1&-\omega&0&0&-a_2&0&0\\
0&0&0&0&0&\omega&a_3&0\\
0&0&0&0&-\omega&0&0&a_3\\
2a_2&0&0&0&-2a_2-k&0&0&\omega\\
0&-a_2&0&0&0&a_2-k&-\omega&0
\end{array}
\right],
\end{align}

\begin{align}\label{eqmz}
{\cal M}_{z}=\left[\begin{array}{cccc}
0&1&0&0\\
-a_1&0&-a_2&0\\
0&0&0&a_3\\
-a_2&0&a_2-1&0
\end{array}
\right],
\end{align}
where
\begin{subequations}\label{abc}
\begin{align}
a_1&=\frac{\delta(1+3Z\delta+3Z\delta^2+Z\delta^3)}{(Z-1)(1+\delta)(1+\delta+2Z\delta^2+Z\delta^3)},\\
a_2&=\frac{\delta}{(1+\delta)(1+\delta+2Z\delta^2+Z\delta^3)},\\
a_3&=\frac{1}{Z-1}.
\end{align}
\end{subequations}
The matrix ${\cal M}_{xy}$ acts on the eight-dimensional vector $\{{\bm\alpha}_1,{\bm\beta}_1,{\bm\alpha}_2,{\bm\beta}_2\}$, while the matrix ${\cal M}_z$ acts on the four-dimensional vector $\{\zeta_1,\sigma_1,\zeta_2,\sigma_2\}$. The eigenvalues of these matrices determine the frequencies of normal modes.
\begin{figure}
\centering
\vspace{0.5cm}
\includegraphics[scale=0.6]{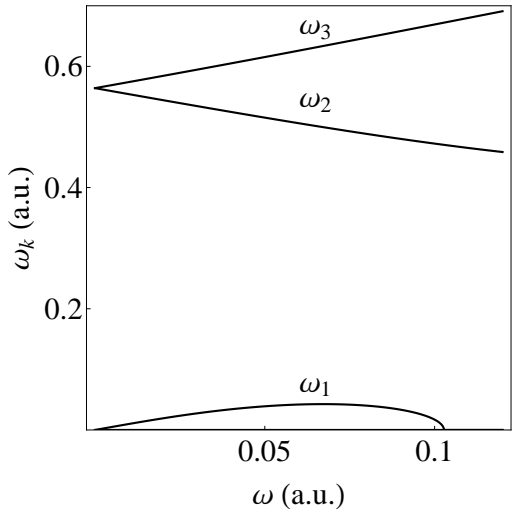}
\caption{Frequencies of the small oscillations in the $xy$ plane as functions of the rotational frequency $\omega$ for magnesium.}\label{fig3}
\end{figure}

\begin{figure}
\centering
\vspace{0.5cm}
\includegraphics[scale=0.6]{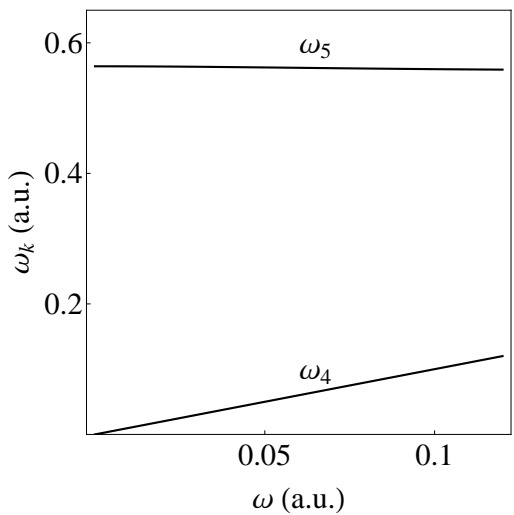}
\caption{Frequencies of the small oscillations in the $z$ direction as functions of the rotational frequency $\omega$ for magnesium.}\label{fig4}
\end{figure}

The matrix ${\cal M}_{xy}$ gives four pairs of frequencies---the members of each pair differ only in sign. One pair consists of zeroes---a result of the broken rotational symmetry. By analogy with field theory, the motion corresponding to the zero frequency may be called the Goldstone mode. The positive partners of the remaining frequencies are plotted in Fig.~\ref{fig3} as functions of the rotational frequency $\omega$ in atomic units. The smallest eigenfrequency $\omega_1$ vanishes at $\omega=$0.1026 a.u. and for larger values of $\omega$ it becomes complex. This sets the stability limit of the equilibrium state. The maximal value of $\delta$ is 0.022. In the stability region, the value of the classical angular momentum $M_z$ changes from $2.8\hbar$ to $\infty$.

The matrix ${\cal M}_z$ gives two pairs of frequencies. The two positive frequencies are plotted in Fig.~\ref{fig4}. In the limit, when $\omega\to 0$ the nonvanishing frequencies $\omega_2,\,\omega_3$ and $\omega_5$ take on the value $\sqrt{k/m_e(Z-1)}$. This frequency can be viewed as the characteristic frequency of core oscillations. For magnesium this value is 0.5637 a.u.

\section{Classical orbits}

\begin{figure}
\centering
\vspace{0.5cm}
\includegraphics[scale=0.7]{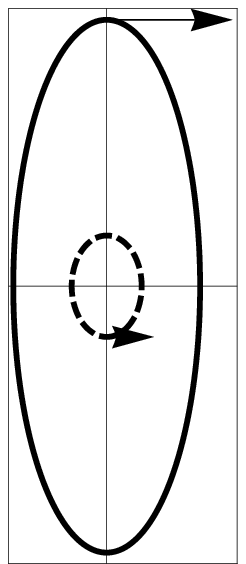}
\caption{Oscillations of the core (dashed line) and the electron in the $xy$ plane around their equilibrium positions for magnesium at $\omega=0.06$ a.u. corresponding to the slow mode, $\omega_1=0.041$ a.u. The size of the core orbit was increased tenfold to make it visible. In the slow mode the core and the electron move in the same directions along the $x$ axis but in the opposite directions along the $y$ axis.}\label{fig5}
\end{figure}

\begin{figure}
\centering
\vspace{0.5cm}
\includegraphics[scale=0.7]{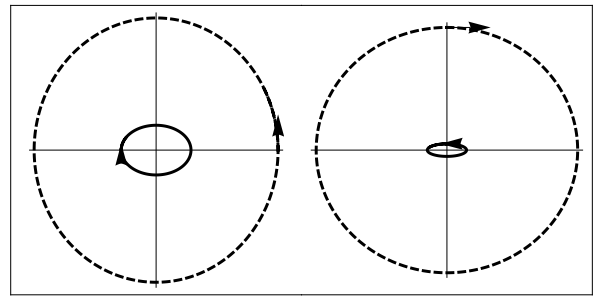}
\caption{Oscillations of the core (dashed line) and the electron around their equilibrium positions for magnesium at $\omega=0.06$ a.u. corresponding to the two fast modes, $\omega_2=0.506$ and $\omega_3=0.626$ in a.u. for the same parameters as in Fig.~\ref{fig5}. In all figures the norm of the amplitude is the same as in Fig.~\ref{fig5}. In both modes the core and the electron move in opposite directions along the $x$ axis. The modes differ in the motion along the $y$ axis. In order to keep the same overall size as in Fig.~\ref{fig5} (assuming the same norm of the amplitude), the overall size was scaled up fivefold.}\label{fig6}
\end{figure}

We begin with the description of small oscillations corresponding to different eigenmodes. To a good approximation, these small oscillations are harmonic---the orbits are elliptical. In Figs.~\ref{fig5} and \ref{fig6} we plotted small oscillations in the $xy$ plane of the core and of the electron around their equilibrium positions for $\omega=0.048$ a.u. They were obtained by calculating numerically the three eigenvectors of the matrix ${\cal M}_{xy}$ corresponding to the eigenfrequencies $\omega_1,\,\omega_2$, and $\omega_3$. In all three modes the motions of the electron and the core are {\em strongly correlated}. In the slow mode (Fig.~\ref{fig5}), corresponding to $\omega_1$, the amplitude of the electron oscillations is much larger than the amplitude of the core oscillations while for the fast modes (Fig.~\ref{fig6}) the amplitudes of the core oscillations dominate.

\begin{figure}
\centering
\includegraphics[scale=0.5]{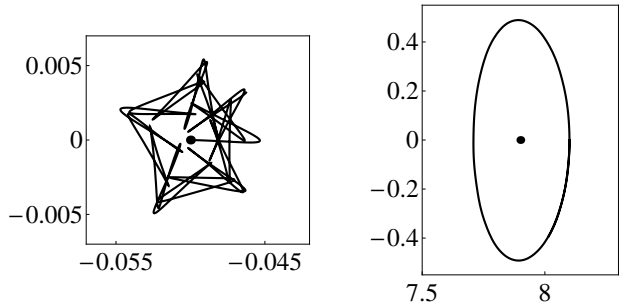}
\caption{The trajectories in the $xy$ plane of the core (left) and the electron (right) in the regime of small oscillations when all modes are excited. The parameter $\omega=0.048$ corresponds to total angular momentum $3\hbar$. At the start the electron was displaced to the right from the equilibrium position (marked by the dot) by 0.2 a.u., while the core started at the equilibrium position. The time lapse is equal to the period of electron as it covers the elliptic orbit. }\label{fig7}
\end{figure}

\begin{figure}
\centering
\includegraphics[scale=0.5]{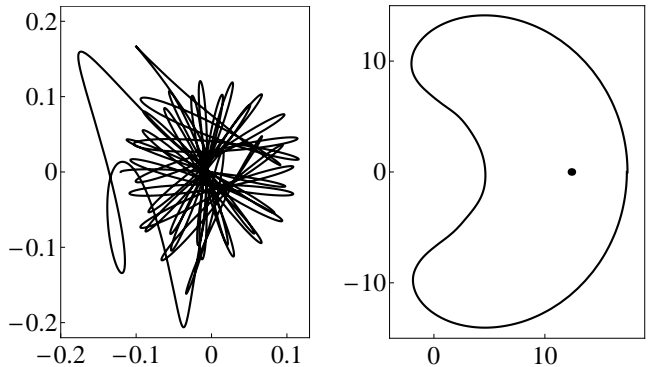}
\caption{The trajectories in the $xy$ plane of the core (left) and the electron (right) during one cycle of the periodic electron motion, outside of the regime of small oscillations. The banana-shaped orbit of the electron shows that the system is fairly robust. The electron was displaced from its equilibrium position (marked by a dot) by as much as 5 a.u. in the horizontal direction and was given the initial velocity 0.12 a.u. in the direction of rotation. The core was initially displaced from its equilibrium position by 0.1 a.u. The total angular momentum of the system is $3\hbar$. The nucleus is in the center of the coordinate system.}\label{fig8}
\end{figure}

Classical orbits of the core and the electron without the approximation of small oscillations can be easily obtained from the numerical solutions of the full equations of motion [Eqs.~(\ref{eqm1})--(\ref{eqm4})]. These solutions show the robustness of our system far beyond the regime of small oscillations. For a random choice of the initial conditions all the modes become excited, the core oscillates rapidly (see Fig.~\ref{fig7}) but the electron follows similar smooth trajectories as in the regime of small oscillations. When the departure from equilibrium is large, the electron trajectory is still regular and stable but it acquires a banana-like shape (see Fig.~\ref{fig8}).

\section{Quantum mechanics of a deformed atom}

In the description of a deformed atom in quantum mechanics we shall start from the same model Hamiltonian (\ref{hamr}) in the rotating frame. Unfortunately, we cannot follow the same path as in the description of the Trojan states \cite{bbke,bbbb} and use the approximation of small oscillations. This is due to the fact that the characteristic frequencies of the electron motion $\omega_1$ and $\omega_4$, as shown in Figs.~\ref{fig3} and \ref{fig4} are tiny. Hence the extension of the electronic Gaussian wave function will exceed the range of small oscillations. Therefore, to find the quantum signature of the lateral localization of the electron vis-a-vis the position of the core we must use a different approach. We shall approximate the interaction potential between the electron and the core by the first two terms of the expansion in the ratio $|{\bm R}|/|{\bm r}|$,
\begin{align}\label{app}
\frac{Z-1}{|{\bm r}-{\bm R}|}\approx\frac{Z-1}{|{\bm r}|}+(Z-1)\frac{{\bm R}\!\cdot\!{\bm r}}{|{\bm r}|^3}.
\end{align}
This approximation is justified, since we aim at the study of electron states localized far from the core and it is also well supported by the classical theory. In this approximation, the Hamiltonian reads:
\begin{align}\label{sh}
H&=-\frac{1}{2}\Delta_r-\frac{1}{|{\bm r}|}+i{\bm\omega}\!\cdot\!({\bm r}\times{\bm\nabla}_r)+(Z-1)\frac{{\bm R}\!\cdot\!{\bm r}}{|{\bm r}|^3}\nonumber\\
&-\frac{1}{2(Z-1)}\Delta_R+\frac{k}{2}{\bm R}^2+i{\bm\omega}\!\cdot\!({\bm R}\times{\bm\nabla}_R).
\end{align}
We took the limit of an infinitely heavy nucleus and we used atomic units. Even after these simplifications, an exact solution of the quantum problem is a hopeless task because the variables cannot be separated. However, we can simplify the problem following the hints provided by the classical theory.

The part of this Hamiltonian involving all the terms describing the core has easily calculable eigenstates. The lowest energy state is described by a shifted Gaussian.
\begin{align}\label{gauss}
\phi({\bm R},{\bm r}) = N\exp\left[-\frac{\gamma}{2}\left({\bm R}+{\bm g}\right)^2+i(Z-1){\bm\omega}\!\cdot\!({\bm R}\times{\bm g})\right],
\end{align}
where
\begin{align}\label{defs}
\gamma &=\sqrt{k(Z-1)},\\
{\bm g}&=\frac{1}{\omega^2|{\bm r}|^3}\left\{\frac{x}{\kappa-1},\frac{y}{\kappa-1},\frac{z}{\kappa}\right\},\\
\kappa &=\frac{k}{(Z-1)\omega^2},
\end{align}
and $N=(\gamma/\pi)^{3/2}$ is the normalization factor. As in the classical case, we have chosen the $z$ axis along ${\bm\omega}$. The core wave function still depends on the electron coordinates through the vector ${\bm g}$. Since the motion of the electron is slow as compared with the fast oscillations of the core, as shown in Fig.~\ref{fig7}, we may approximate the wave function $\Psi({\bm R},{\bm r})$ of the whole system by a Born-Oppenheimer--like product.
\begin{align}\label{bo}
\Psi({\bm R},{\bm r})=\phi({\bm R},{\bm r})\psi({\bm r}).
\end{align}
Within this approximation we can confirm our classical picture of the deformed atom in which the electron and the core are localized on the opposite sides of the nucleus since the expectation value of the scalar product ${\bm R}\!\cdot\!{\bm r}$ is strictly negative,
\begin{align}\label{cos}
\int\!d^3Rd^3r|\phi({\bm R},{\bm r})\psi({\bm r})|^2{\bm R}\!\cdot\!{\bm r}
=-\int\!d^3r|\psi({\bm r})|^2{\bm g}\!\cdot\!{\bm r}.
\end{align}
We carried out here the Gaussian integration over ${\bm R}$, after a change of variables ${\bm R}\to {\bm R}-{\bm g}$.

\begin{figure}
\centering
\vspace{0.5cm}
\includegraphics[scale=0.5]{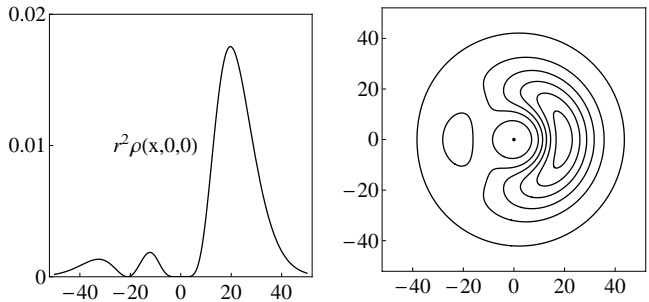}\caption{Plots of the electron density along the $x$ axis (left) and of the density in the $xy$ plane (right) for the superposition of $n=5$ and $n=4$ states. The dot indicates the position of the nucleus. The interpretation of contour lines is made possible with the plot of the effective density $r^2\rho$ along the $x$ axis in the equatorial plane. The coordinates $x$ and $y$ are measured in atomic units}\label{fig9}
\end{figure}

\begin{figure}
\centering
\vspace{0.5cm}
\includegraphics[scale=0.5]{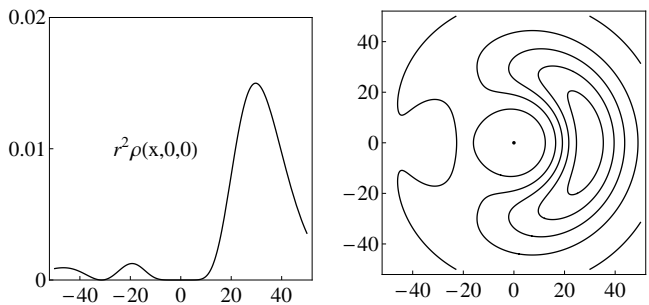}\caption{The same plots as in Fig.~\ref{fig9} but for the superposition of $n=6$ and $n=5$ states.}\label{fig10}
\end{figure}

Next, we will determine the shape of the electron wave function while the position of the core ${\bm R}$ is kept fixed at its classical equilibrium value ${\bm R}=(-x_2,0,0)$. This assumption is based on the classical analysis which showed that the core undergoes rapid oscillations in the close vicinity of the equilibrium position while the outer electron moves much more slowly (see Fig.~\ref{fig7}). Under these assumptions the Schr\"odinger equation for the electronic wave function $\psi({\bm r})$ is
\begin{align}\label{sh1}
\left(H_0+H_1\right)\psi({\bm r})=E\psi({\bm r}),
\end{align}
where
\begin{align}\label{sh2}
H_0&=-\frac{1}{2}\Delta-\frac{1}{|{\bm r}|}+i{\bm\omega}\!\cdot\!({\bm r}\times{\bm\nabla}),\\
H_1&=(Z-1)\frac{{\bm R}\!\cdot\!{\bm r}}{|{\bm r}|^3}.
\end{align}
In the region of large $r$, where we expect electron localization, the dipole term $H_1$ is very small and can be treated perturbatively. The eigenfunctions $\psi_{nlm}(r,\theta,\phi)$ of $H_0$ are well known. The only possibility for a small perturbation to produce a large modification of the wave functions $\psi({\bm r})$ is the occurrence of a resonance. Due to the presence of the rotational term $-\omega M_z$ in $H_0$, there is a possibility of such a resonance. This is a typical case of level crossing and it happens when the rotational contribution $\omega M_z$ to the Hamiltonian matches the difference between the hydrogenic energy levels $E_n=-1/2n^2$ in the rotating frame,
\begin{align}\label{res}
E_{n_1}-\omega m_1=E_{n_2}-\omega m_2.
\end{align}
In such a case we have to deal with the degenerate perturbation theory. The Hamiltonian $H_1$ has nonvanishing matrix elements between pairs of states that satisfy the dipole selection rules $\Delta l=\pm 1, \Delta m=\pm 1$. For definiteness, let us choose $n_1>n_2$. Without loss of generality, we may assume that $\Delta m=m_1-m_2=1$ and we obtain the resonance value of the rotational frequency,
\begin{align}\label{res1}
\omega_{n_1n_2}=E_{n_1}-E_{n_2}.
\end{align}
The opposite sign of $\Delta m$ would only give the reversed direction of rotation. According to our classical analysis, the deformation will be the largest when the electron stays near the equatorial plane and is sufficiently far from the nucleus. This means that $l$ and $m$ should be the largest possible. This observation goes very well with the fact that matrix elements of $H_1$ between the hydrogenic wave functions $\psi_{nlm}(r,\theta,\phi)$ are maximal for the largest values of $l$ and $m$. Last but not least, we must stay within the limits of our version of the Born-Oppenheimer approximation and of the dipole approximation---both require the outer electron to be far from the core. On the other hand, the interaction between the core and the electrons weakens with the increasing distance between them.

We shall build the quantum state of the deformed atom from Coulomb states with maximal values of the quantum numbers $l$ and $m$, for a given $n$. Under this assumption, the subspace of states with the same eigenvalue of the unperturbed Hamiltonian, for the rotational frequency $\omega$ satisfying Eq.~(\ref{res1}), is two-dimensional. For each $n$, this subspace is built on two hydrogenic states: $|n,\,n-1,\,n-1\rangle$ and $|n-1,\,n-2,\,n-2\rangle$. The degeneracy is removed by the interaction term represented by the matrix
\begin{align}\label{dm}
\left(\begin{array}{cc}
0&d_n\\
d_n&0
\end{array}\right),
\end{align}
where $d_n$ is the matrix element of the interaction Hamiltonian $H_1$ taken between the hydrogenic wave functions,
\begin{align}\label{deg}
&d_n=(Z-1)\langle n,n-1,n-1|\frac{{\bm R}\!\cdot\!\bm r}{|{\bm r}|^3}|n-1,n-2,n-2\rangle\nonumber\\
&=-x_2(Z-1)\langle n,n-1,n-1|\frac{x}{|{\bm r}|^3}|n-1,n-2,n-2\rangle.
\end{align}
The normalized eigenvectors of the matrix (\ref{dm}) have the form:
\begin{align}\label{ev}
|n\pm\rangle=\frac{1}{\sqrt{2}}\left(|n,n-1,n-1\rangle\pm|n-1,n-2,n-2\rangle\right).
\end{align}
These states correspond to the values of the quasienergy $E_n-(n-1)\omega\pm d_n$.

Superpositions of states with different values of the magnetic quantum number {\em always} lead to wave packets of the electron density on the orbit. However, in general, such wave packets will not be stable. In our model, due to the interaction with deformed core, the energy difference stabilizes the deformed states.

The superposition [Eq.~(\ref{ev})] with the plus sign, gives the electron density localized on the same side of the nucleus relative to the average position of the core described by $\bm R$. This state has a higher quasienergy ($d_n>0)$. The superposition with the minus sign leads to the localization of the electron on the opposite side of the nucleus. Therefore, it corresponds to the lower quasienergy. This quantum state, together with Gaussian state of the core $\phi({\bm R},{\bm r})$, is a counterpart of classical localized orbits of the electron and the core---it gives a quantum description of a deformed atom.

To illustrate these phenomena we choose the following superpositions, including their time-dependent phase factors
\begin{subequations}\label{56}
\begin{align}
|5-\rangle=\exp[-i(E_5-4\omega_{54}-d_5)t] \frac{1}{\sqrt{2}}\left(|5,4,4\rangle-|4,3,3\rangle\right),\label{54}\\
|6-\rangle=\exp[-i(E_6-5\omega_{65}-d_6)t] \frac{1}{\sqrt{2}}\left(|6,5,5\rangle-|5,4,4\rangle\right).\label{65}
\end{align}
\end{subequations}
The electron densities are depicted in Figs.~\ref{fig9} and \ref{fig10}. We see that the localization weakens with the increase of the principal quantum number. This was to be expected since a larger $n$ means a weaker interaction with the core.

In what follows, we shall focus on the state $|5-\rangle$. The matrix element in Eq.~(\ref{deg}) for $n=5$ is equal to $-0.001$. According to (\ref{res1}), in this case the value of $\omega=\omega_{54}$ is $1/32-1/50$ and $x_2=0.0079$. Therefore, the quasienergy for this state is decreased by $d_5=9\times 10^{-5}$.

It is instructive to transform the stationary state in the rotating frame to the (inertial) laboratory frame. This transformation is accomplished simply by multiplying every eigenstate of $M_z$ belonging to the eigenvalue $\hbar m$ by the phase factor $\exp(-i\omega m t)$. This means that the state vector (\ref{54}) transforms into
\begin{align}\label{5}
|5-\rangle &=e^{-i(E_5-4\omega-d_5) t}\frac{1}{\sqrt{2}}\left(e^{-i4\omega t}|5,4,4\rangle-e^{-i3\omega t}|4,3,3\rangle\right)\nonumber\\
&=e^{-i(E_5-d_5) t}\frac{1}{\sqrt{2}}\left(|5,4,4\rangle-e^{i\omega t}|4,3,3\rangle\right).
\end{align}
The electron probability density in this state is given by the formula:
\begin{align}\label{rhot}
\rho(r,\theta,\phi,t)&=\frac{1}{2}\left(|\psi_{5,4,4}(r,\theta)|^2
+|\psi_{4,3,3}(r,\theta)|^2\right)\nonumber\\
&-\psi_{5,4,4}(r,\theta)\psi_{4,3,3}(r,\theta)\cos(\varphi-\omega t),
\end{align}
where we denoted by $\psi_{nlm}(r,\theta)$ the hydrogenic wave function without the phase factors $\exp(im\varphi)$ and we have taken into account the fact that these functions for the maximal values of $l$ and $m$ are real. Thus, as expected, the superposition (\ref{54}) transformed to the laboratory frame describes an electron wave packet revolving around the nucleus with the frequency (\ref{res1}).

\section{Radiative decays and the rotational frequency shift}

The quantum state $|5-\rangle$ of the deformed atom has only two decay channels with the emission of dipole radiation. The selection rules and the quasienergy conservation allow the transitions to the states $|5+\rangle$ and $|3,2,2\rangle$. In the first case, the radiated photon has the quasienergy $-2 d_5$ and it is linearly polarized along the $y$ direction. The dipole matrix element $\langle 5+|y|5-\rangle$ is equal to $-8.35i$. In the second case, the quasienergy of the photon is $2E_4-E_5-E_3-d_5$ and the radiated photon is emitted with circular polarization. The dipole matrix element $\langle 3,2,2|(x-iy|5-\rangle$ is equal to $\sqrt{2}4.74$. These results must be now transformed to the laboratory frame.

Deformed states of the atom offer a unique opportunity to study the effects of the rotational frequency shift predicted some time ago \cite{bbbb2}. These effects up to now have been seen only in molecules \cite{mhs}. Rotational frequency shift is an analog of the Doppler shift. To account for the Doppler shift, when comparing the energies of a photon with momentum ${\hbar\bm k}$ measured in two reference frames moving with respect to each other with velocity ${\bm v}$, one has to add the scalar product ${\bm v}\!\cdot\!{\bm k}$.  To account for the rotational frequency shift, when comparing the energies of a photon with angular momentum ${\bm M}$ measured in two reference frames rotating with respect to each other with the angular velocity ${\bm\omega}$, one has to add the scalar product ${\bm\omega}\!\cdot\!{\bm M}$ of the angular velocity and photon angular momentum. The rotational frequency shift does not lead to any observable phenomena for a state that is rotationally symmetric since there is no way of telling whether a quantum system in such a state is rotating or not.

The transformation to the laboratory frame does not affect the transition matrix elements but the energy of the emitted photons acquires an additional term $\omega=E_5-E_4$. Thus, the spontaneous decay of our state $|5-\rangle$ leads to the emission of a linearly polarized photon with the energy $E_5-E_4-2d_5$ or a circularly polarized photon with the energy $E_4-E_3-d_5$. The transition rates are practically the same as for standard Rydberg states with maximal values of $l$ and $m$.

\section{Conclusions}

With the use of a simple model of an isolated multielectron atom we predicted the existence of stable states describing a deformed atom. In such states, one excited electron is localized as a wave packet on a circular Rydberg orbit. The core made of the remaining electrons is polarized due to the interaction between the electron and the core. The deformation requires breaking of the rotational symmetry. In classical theory, we were able to give exact description of the atomic deformation and we proved the linear stability of deformed states. In quantum theory, we had to resort to a perturbative treatment of the interaction between the core and the electron. Localized states of the excited electron are special superpositions of two circular Rydberg states. In the construction of such stable localized states, the essential role was played by the phenomenon of level crossing in the frame rotating with the electron.

We cannot give a prescription on how to produce the special superpositions describing a localized electron. However, the calculated redshifts in spontaneous decays of those states could be used as their signature. Unfortunately, our model is so crude that its predictions cannot be precise but we hope that they are of the right order of magnitude.

\acknowledgments

We acknowledge support from the Polish Ministry of Science and Higher Education under a grant for the years 2008--2010.

\end{document}